\documentclass[a4paper,runningheads]{llncs}
\usepackage{amsmath}   
\usepackage{amssymb}
\usepackage{url}

\usepackage{latexsym}           

\usepackage{epsfig,color}             
\usepackage{epic}
\usepackage{eepic}
\usepackage{ecltree}

\usepackage{ifthen}             
\usepackage{xspace}             
\catcode`!=11
\def\!ifnextchar#1#2#3{%
  \let\!testchar=#1%
  \def\!first{#2}%
  \def\!second{#3}%
  \futurelet\!nextchar\!testnext}
\def\!testnext{%
  \ifx \!nextchar \!spacetoken 
    \let\!next=\!skipspacetestagain
  \else
    \ifx \!nextchar \!testchar
      \let\!next=\!first
    \else 
      \let\!next=\!second 
    \fi 
  \fi
  \!next}

\usepackage{makeidx}

\def\!define#1[#2]{\textsl{\textbf{{#1}}}\index{#2}}
\catcode`!=12

\newcommand{\comment}[1]{}

\DeclareSymbolFont{AMSb}{U}{msb}{m}{n}
\DeclareSymbolFontAlphabet{\mathbb}{AMSb}

\newcommand{\un}{\sqcup}                     

\def\acurule(#1,#2,#3,#4){#2.#3\stackrel{#1(#2)}{\longrightarrow} #4}

\def\acbrule(#1,#2,#3){#1\un #2\ra #3}

\def\move(#1){\rp {#1}}           
\def\nmove(#1,#2){\rp {#1}^{#2}}  

\def\th{^{\mathrm{th}}\kern+.29em}        
\def\card(#1){\mathsf{Card}(#1)}         
\def\sol(#1){\mathsf{SOL}(#1)}           
\def\cond(#1){\mathsf{COND}(#1)}         

\renewcommand{\int}{\mbox{\sf int}}

\def\A{{\cal A}}

\def\C{{\cal C}}

\newcommand{\ra}{\rightarrow}

\newcommand{\F} {{\cal F}}                    
\def\TGF(#1){T(#1)}                           
\def\TGFX(#1,#2){T(#1,#2)}                    
\newcommand{\TF} {\TGF(\F)}                   
\def\CTXG(#1){{\cal C}(#1)}                     
\def\CTXGN(#1,#2){{\cal C}^#2(#1)}                    
\def\CTX{\CTXG(\F)}                             

\let\ouvsub\{
\let\fermsub\}
\def\flechsub{\negthickspace\leftarrow\negmedspace}
\def\subst(#1,#2,#3){\ifthenelse{\equal{#3}{1}}
  {\ouvsub#1\flechsub#2\fermsub}
  {\ifthenelse{\equal{#3}{2}}
    {\ouvsub#1_1\flechsub#2_1,#1_#3\flechsub#2_#3\fermsub}
    {\ouvsub#1_1\flechsub#2_1,\ldots,#1_#3\flechsub#2_#3\fermsub}}
  }

\def\seq(#1,#2){\ifthenelse{\equal{#2}{2}}%
                  {{#1}_1,{#1}_{#2}}%
                  {{#1}_1,\ldots,{#1}_{#2}}%
               }
\def\seqx(#1,#2,#3){\ifthenelse{\equal{#3}{2}}%
                  {{#1}_1({#2}_1),{#1}_{#3}({#2}_{#3})}%
                  {{#1}_1({#2}_1),\ldots,{#1}_{#3}({#2}_{#3})}%
                   }
\def\tuple(#1,#2){(\seq(#1,#2))}                

\def\fta(#1,#2,#3,#4){(#1,#2,#3,#4)}     
\def\aaa{\fta(Q,\F,Q_f,\Delta)}                  
\def\rulea(#1,#2,#3,#4,#5){#1(\seqx(#2,#3,#4))\ra #5(#1\tuple(#3,#4))}
\def\urule(#1,#2){#1\ra #2(#1)}

\def\grule(#1,#2,#3,#4){#1(\seq(#2,#3))\ra #4}
\def\stdgrule{\grule(f,q,n,q)}                  
\def\gurule(#1,#2){#1\ra #2}

\def\ada{\fta(Q,\F,I,\Delta)}                  
\def\ruled(#1,#2,#3,#4,#5){#1(#2\tuple(#3,#4))\ra #5(\seqx(#1,#3,#4))}
\def\uruled(#1,#2){#1(#2)\ra #2}
\def\stduruled{\uruled(q,a)}       

\def\linefill{$\m@th\mathord-\mkern-6mu%
  \cleaders\hbox{$\mkern-2mu\mathord-\mkern-2mu$}\hfill
  \mkern-6mu\mathord-$}

\newcommand{\eqstep}[2]                                 
{
~\mathop{\mbox{\leftarrowfill$\!\!$\rightarrowfill}}
\limits^{~{#1}~}_{~{#2}~}~
}

\newcommand{\eqstepm}[3]                                
{
~\mathop{\mbox
{\leftarrowfill$\!\scriptscriptstyle{#1}\!$\rightarrowfill}}
\limits^{~{#2}~}_{~{#3}~}~
}

\newcommand{\lrstepm}[3]                                
{
~\mathop{\mbox
{\linefill$\!\scriptscriptstyle{#1}\!$\rightarrowfill}}
\limits^{~{#2}~}_{~{#3}~}~
}

\newcommand{\rlstepm}[3]                                
{
~\mathop{\mbox
{\leftarrowfill$\!\scriptscriptstyle{#1}\!$\linefill}}
\limits^{~{#2}~}_{~{#3}~}~
}

\newcommand{\rlstep}[2]                                 
{
~\mathop{\mbox{\leftarrowfill}}
\limits^{~{#1}~}_{~{#2}~}~
}

\newcommand{\lrstep}[2]                                 
{
~\mathop{\mbox{\rightarrowfill}}
\limits^{~{#1}~}_{~{#2}~}~
}
\def\lrproof#1{\lrstep{*}{#1}}                          

\newcommand{\rp}[1]{\mathop{\rightarrow}_{#1}}                  

\newcommand{\encomp}{\mbox{$\:\:\:\mathop{\raisebox{0.24ex}[0mm][0mm]{$\cdot$}\!\!\!\unlhd}$}}

\newcommand{\subsm}[1]
{\mathop{\raisebox{0.24ex}[0mm][0mm]{$\cdot$}\!\!\!\!\geq_{#1}}}
\newcommand{\enco}
{\encomp}

\renewcommand{\lrproof}[1]{\mathop{\rightarrow_{#1}^{*}}}
\renewcommand{\lrstep}[2]{\mathop{\rightarrow_{#2}}}

\newenvironment{sketchofproof}{\par\smallskip\par\noindent\emph{Sketch
    of proof}}{\qed }

\def\F{\mathcal F}
\def\C{\mathcal C}

\def\T{\mathcal T}
\begin{document}
\bibliographystyle{alpha}
\title{Residual Finite Tree Automata\thanks{This research was
  partially supported by ``TACT-TIC'' r\'egion Nord-Pas-de-Calais ---
  FEDER and the MOSTRARE  INRIA project} }

\author{J. Carme \and  R. Gilleron \and A. Lemay \and A. Terlutte \and
  M. Tommasi}

\institute{Grappa -- EA 3588 -- Lille 3 University\\ \url{http://www.grappa.univ-lille3.fr}}

\maketitle             

\begin{abstract}
  Tree automata based algorithms are essential in many fields in
  computer science such as verification, specification, program
  analysis. They become also essential for databases with the
  development of standards such as XML. In this paper, we define new
  classes of non deterministic tree automata, namely residual finite
  tree automata (RFTA). In the bottom-up case, we obtain a new
  characterization of regular tree languages. In the top-down case, we
  obtain a subclass of regular tree languages which contains the class
  of languages recognized by deterministic top-down tree automata.
  RFTA also come with the property of existence of canonical non
  deterministic tree automata.

\end{abstract}


\section{Introduction}

The study of tree automata has a long history in computer science; see
the survey of Thatcher~\cite{Thatcher73}, and the texts of
F.~G\'{e}cseg and M.~Steinby~\cite{GecsegSteinby84,GecsegSteinby96}, 
and of the TATA
group~\cite{tata97}. With the advent of tree-based metalanguages (SGML
and XML) for document grammars, new developments on tree automata
formalisms and tree automata based algorithms have been
done~\cite{murata01taxonomy,Neven02}. Also, because of the tree
structure of documents, learning algorithms for tree languages have
been defined for the tasks of information extraction and information
retrieval~\cite{Fernau02,GoldmanKwek02,xwrap}. We are currently
involved in a research project dealing with information extraction
systems from semi-structured data. One objective is the definition of
classes of tree automata satisfying two properties: there are
efficient algorithms for membership and matching, and there are
efficient learning algorithms for the corresponding classes of tree
languages.

In the present paper, we only consider finite ranked trees. 
There are
bottom-up (also known as frontier to root) tree automata and top-down
(also known as root to frontier) tree automata. The top-down version
is particularly relevant for some implementations because important
properties such as membership\footnote{given a tree automaton $A$,
  decide whether an input tree is accepted by $A$.} can be solved
without handling the whole input tree into memory.  There are also
deterministic tree automata and non-deterministic tree automata.
Determinism is important to reach efficiency for membership and other
decision properties. It is known that non-deterministic top-down,
non-deterministic bottom-up, and deterministic bottom-up tree automata
are equally expressive and define regular tree languages. But there is
a tradeoff between efficiency and expressiveness because some regular
(and even finite) tree languages are not recognized by deterministic
top-down tree automata. Moreover, the size of a deterministic
bottom-up tree automaton can be exponentially larger than the size of
a non-deterministic one recognizing the same tree language. This
drawback can be dramatic when the purpose is to build tree automata.
This is for instance the case in the problem of tree pattern matching
and in machine learning problems like grammatical inference.  

The process of learning finite state machines from data is referred as
grammatical inference. The first theoretical foundations were given by
Gold~\cite{Gold67} and first applications were designed in the field
of pattern recognition. Grammatical inference mostly focused on
learning string languages but recent works are concerned with learning
tree languages~\cite{Sakakibara90,Fernau02,GoldmanKwek02}. In most
works, the target tree language is represented by a deterministic
bottom-up tree automaton. This is problematic because the time
complexity of the learning algorithm depends on the size of the target
automaton. Therefore, again it is crucial to define learning
algorithms for non-deterministic tree automata. The reader should note
that tree patterns~\cite{GoldmanKwek02} satisfy this property.

Therefore the aim of this article is to define non-deterministic tree
automata corresponding to sufficiently expressive classes of tree
languages and having nice properties from the algorithmic viewpoint
and from the grammatical inference viewpoint. For this aim, we extend
previous works from the string case~\cite{DenisLemayTerlutte2002a} to
the tree case and we define residual finite state automata (RFTA). The
reader should note that learning algorithms for residual finite string
automata have been
defined~\cite{DenisLemayTerlutte2001c,DenisLemayTerlutte2002b}.

In Section~\ref{sec:bottomup}, we study the bottom-up case. We define
the residual language of a language $L$ w.r.t a ground term $t$ as
the set of contexts $c$ such that $c[t]$ is a term in $L$.  We define
bottom-up residual tree automata as automata whose states correspond
to residual languages. Bottom-up residual tree automata are
non-deterministic and recognize regular tree languages. We prove that
every regular tree language is recognized by a unique canonical
bottom-up residual tree automaton, minimal according to the number of
states.  We give an example of regular tree languages for which the
size of the deterministic bottom-up tree automata grows exponentially
with respect to the size of the canonical bottom-up residual tree
automata.

In Section~\ref{sec:topdown}, we study the top-down case. We define
the residual language of a language $L$ w.r.t a context $c$ as the
set of ground terms $t$ such that $c[t]$ is a term in $L$.  We define
top-down residual tree automata as automata whose states correspond to
residual languages. Top-down residual tree automata are
non-deterministic tree automata. Interestingly, the class of languages
recognized by top-down residual tree automata is strictly included in
the class of regular tree languages and strictly contains the class of
languages recognized by deterministic top-down tree automata. We also
prove that every tree language in this family is recognized by a
unique canonical top-down residual tree automaton; this automaton is
minimal according to the number of states.

The definition of residual finite state automata comes with new
decision problems. All of them rely on properties of residual
languages. It is proved that all residual languages of a given tree
language $L$ can be built in both top-down and bottom-up cases.  From
these constructions we obtain positive answers to decision problems
like 'decide whether an automaton is a (canonical) RFTA'. The exact
complexity bounds are not given but we conjecture that are identical
than in the string case.

The present work is connected with the paper by Nivat and
Podelski~\cite{NivatPodelski97}. They consider a monoid framework,
whose elements are called pointed trees (contexts in our terminology,
special trees in~\cite{Thomas84}),
to define tree automata. They define a Nerode congruence in the
bottom-up case and in the top-down case. Their work leads to the
generalization of the notion of deterministic to l-r-deterministic
(context-deterministic in our terminology) for top-down tree automata.
They have a minimization procedure for this class of automata. It
should be noted that the class of languages recognized by
context-deterministic tree automata (also called homogeneous tree
languages) is strictly included in the class of languages recognized
by residual top-down tree automata.

%
%
\section{Preliminaries}

We assume that the reader is familiar with basic knowledge about tree
automata. We follow the notations defined in TATA~\cite{tata97}.

A ranked alphabet is a couple $(\F,\emph{Arity})$ where $\F$ is a
finite set and $\emph{Arity}$ is a mapping from $\F$ into $\mathbb N$.
The set of symbols of arity $p$ is denoted by $\F_p$. Elements of
arity $0$, $1$, \dots $p$ are respectively called constants, unary,
\dots, $p$-ary symbols.  We assume that $\F$ contains at least one
constant. In the examples, we use parenthesis and commas for a short
declaration of symbols with arity. For instance, $a$ is a constant and
$f(,)$ is a short declaration for a binary symbol $f$.  The set of
\emph{terms} over $\F$ is denoted by $\T(\F)$.  Let $\diamond$ be a
special constant which is not in $\F$. The set of \emph{contexts}
(also known as pointed trees in~\cite{NivatPodelski97} and special
trees in~\cite{Thomas84}), denoted by $\CTX$, is the set of terms
which contains exactly one occurrence of $\diamond$. The expression
$c[\diamond]$ denotes a context, we only write $c$ when there is no
ambiguity.  We denote by $c[t]$ the term obtained from $c[\diamond]$
by replacing $\diamond$ by a term $t$.

A \emph{bottom-up Finite Tree Automaton} ($\uparrow$-FTA) over $\F$ is
a tuple $A=\aaa$ where $Q$ is a finite set of states, $Q_f \subseteq
Q$ is a set of final states, and $\Delta$ is a set of transition rules
of the form $\stdgrule$ where $n\geq 0$, $f \in \F_n$, $q,\seq(q,n)
\in Q$. In this paper, the size of an automaton refers to its size in
number of states, so two automaton which have the same number of
states but different number of rules are considered as having
 the same
size.  When $n=0$ a rule is written $a \rightarrow q$, where $a$ is a
constant.  The \emph{move relation} is written $\lrstep{}{A}$ and
$\lrproof{A}$ is the reflexive and transitive closure of
$\lrstep{}{A}$. A term $t$ reaches a state $q$ if and only if $t
\lrproof{A} q$. A state $q$ \emph{accepts} a context $c$ if and only
if there exists a $q_f \in Q_f$ such that $c[q] \lrproof{A} q_f$.  The
automaton $A$ recognizes a term $t$ if and only if there exists a $q_f
\in Q_f$ such that $t \lrproof{A} q_f$. The language recognized by $A$
is the set of all terms recognized by $A$, and is denoted by $L(A)$.

Two $\uparrow$-FTA are equivalent if they recognize the same tree
language. A $\uparrow$-FTA $A=\aaa$ is \emph{trimmed} if and only if
all its states can be reached by at least one term and accepts at
least one context. A $\uparrow$-FTA is \emph{deterministic}
($\uparrow$-DFTA) if and only if there are no two rules with the same
left-hand side in its set of rules. A tree language is \emph{regular}
if and only if it is recognized by a bottom-up tree automaton. As any
$\uparrow$-FTA can be changed into an equivalent trimmed
$\uparrow$-DFTA, any regular tree language can be recognized by a
trimmed $\uparrow$-DFTA.

Let $L$ be a tree language over a ranked alphabet $\F$ and $t$ a term.
The bottom-up residual language of $L$ relative to a term $t$,
denoted by $t^{-1}L$, is the set of all contexts in $\C(\F)$ such that
$c[t] \in L$:
$$t^{-1} L = \{ c \in \C(\F) \mid c[t] \in L\}.$$
Note that a
bottom-up residual language is a set of contexts, and not a tree
language. The Myhill-Nerode congruence for tree languages can be
defined by two terms $t$ and $t'$ are equivalent if they define the
same residual languages. From the Myhill-Nerode theorem fro tree
languages, we get the following result: a tree language is
recognizable if and only if the number of residual languages is
finite.



A \emph{top-down finite tree automaton} ($\downarrow$-FTA) over $\F$
is a tuple ${\cal A} = \ada$ where $Q$ is a set of states, $I\subseteq
Q$ is a set of initial states, and $\Delta$ is a set of rewrite rules
of the form $q(f)\ra f(\seq(q,n))$ where $n\geq 0$, $f \in \F_n$, $q,
\seq(q,n) \in Q$. Again, if $n=0$ the rule is written $\stduruled$.
The \emph{move relation} is written $\lrstep{}{A}$ and $\lrproof{A}$
is the reflexive and transitive closure of $\lrstep{}{A}$. A state $q$
accepts a term $t$ if and only if $q(t) \lrproof{A} t$. $A$ recognizes
a term $t$ if and only if at least one of its initial states accepts
it. The language recognized by $A$ is the set of all ground terms
recognized by $A$ and is denoted by $L(A)$.

Any regular tree language can be recognized by a $\downarrow$-FTA.
This means that $\downarrow$-FTA and $\uparrow$-FTA have the same
expressive power.  A $\downarrow$-FTA is \emph{deterministic}
($\downarrow$-DFTA) if and only if its set of rules does not contain
two rules with the same left-hand side. Unlike $\uparrow$-DFTA,
$\downarrow$-DFTA are not able to recognize all regular tree languages.

Let $L$ be a tree language over a ranked alphabet $\F$, and $c$ a
context of $\C(\F)$. The top-down residual language of $L$ relative to
$c$, denoted by $c^{-1}L$, is the set of ground terms $t$ such that
$c[t] \in L$:
$$c^{-1} L = \{ t \in \T(\F) \mid c[t] \in L\}.$$

The definition of top-down residual languages comes with an
equivalence relation on contexts. It is worth noting that it does not
define a congruence over terms. Nonetheless, based
on~\cite{NivatPodelski97}, it can be shown that a tree language $L$ is
regular if and only if the number of top-down residual languages
associated with $L$ is finite. In the proof, it is used that the
number top-down residual languages is lower than the number of
bottom-up residual languages.


\section{Bottom-up residual finite tree automata}
\label{sec:bottomup}


In this section, we introduce a new class of bottom-up finite tree
automata, called bottom-up residual finite tree automata
($\uparrow$-RFTA).  This class of automata shares some interesting
properties with both bottom-up deterministic and non-deterministic
finite tree automata which both recognize the class of regular tree
languages.

On the one hand, as $\uparrow$-DFTA, $\uparrow$-RFTA admits a unique
canonical form, based on a correspondence between states and residual
languages, whereas $\uparrow$-FTA does not. On the other hand,
$\uparrow$-RFTA are non-deterministic and can be much smaller in their
canonical form than their deterministic counter-parts.

\subsection{Definition and expressive power of bottom-up residual 
finite tree automata}

First, let us precise the nature of this correspondence, then let us
give the formal definition of $\uparrow$-residual tree automata and
describe their properties.

In order to establish the nature of this correspondence between states
and residual languages, let us introduce the notion of state
languages. The \emph{state language} $C_q$ of a state $q$ is the set
of contexts accepted by the state $q$:

$$C_q = \{ c \in \C(\F) \mid \exists q_f \in Q_f, c[q] \lrproof{A} q_f \}.$$

As shown by the following example, state languages are generally not
residual languages:
\begin{example} \label{ex:statelanguagesnotresiduallanguages}
  Consider the tree language $L=\{f(a_1,b_1),f(a_1,b_2),f(a_2,b_2)\}$
  over $\F=\{f(,),a_1,b_1,a_2,b_2\}$. This language $L$ is recognized
  by the tree automaton $A=\fta(\{ q_1,q_2,q_3,q_4,q_5 \},\F,\{ q_5
  \},\Delta)$ where $\Delta= \{ a_1 \rightarrow q_1, b_1 \rightarrow
  q_2, b_2 \rightarrow q_3, a_2 \rightarrow q_4, a_1 \rightarrow q_4,
  f(q_1,q_2) \rightarrow q_5, f(q_4,q_3) \rightarrow q_5 \}$.
  Residual languages of $L$ are
  $a_1^{-1}L=\{f(\diamond,b_1),f(\diamond,b_2)\}$,
  $b_1^{-1}L=\{f(a_1,\diamond)\}$,
  $b_2^{-1}L=\{f(a_1,\diamond),f(a_2,\diamond)\}$,
  $a_2^{-1}L=\{f(\diamond,b_2)\}$, $f(a_1,b_1)^{-1}L=\{\diamond\}$.
  The state language of $q_1$ is $\{ f( \diamond ,b_1) \}$, which is
  not a residual language.  The tree $a_1$ reaches $q_1$, so each
  context accepted by $q_1$ is an element of the residual language
  $a_1^{-1}L$, which means that $C_{q_1} \subset a_1^{-1}L$. But the
  reverse inclusion is not true because$f(\diamond,b_2)$ is not an
  element of $C_{q_1}$. The reader should note that this situation is
  possible because $A$ is non-deterministic.
\end{example}

In fact, it can be proved (the proof is omitted) that residual
languages are unions of state languages. For any $L$ recognized by a
tree automaton $A$, we have
\begin{equation}
\forall t \in \TF, t^{-1}L = \bigcup_{q \in Q , \; t \lrproof{A} q}
C_q.\label{eq:1}
\end{equation}

As a consequence, if $A$ is deterministic and trimmed, each residual
language is a state language and conversely. 

We can define a new class of non-deterministic automata stating that
each state language must correspond to a residual tree language. We
have seen that residual tree languages are related to the
Myhill-Nerode congruence and we will show that minimization of tree
automata can be extended in the definition of a canonical form for
this class of non-deterministic tree automata.


\begin{definition}
  A bottom-up residual tree automaton ($\uparrow$-RFTA) is a
  $\uparrow$-FTA $A=\aaa$ such that $\forall q \in Q$, $\exists t \in
  \TF$, $C_q = t^{-1}L(A).$
\end{definition}

According to the above definition and previous remarks, it can be
shown that every trimmed $\uparrow$-DFTA is a $\uparrow$-RFTA. As a
consequence, $\uparrow$-RFTA have the same expressive power than
finite tree automata:

\begin{theorem}
  The class of tree languages recognized by $\uparrow$-RFTA is the
  class of regular tree languages.
\end{theorem}

As an advantage of $\uparrow$-RFTA, the number of states of an
$\uparrow$-RFTA can be much smaller than the number of states of any
equivalent $\uparrow$-DFTA:

\begin{proposition} \label{prop:exporftadfta}
  There exists a sequence $(L_n)$ of regular tree languages such that for
  each $L_n$, the size of the smallest $\uparrow$-DFTA which
  recognizes $L_n$ is an exponential function of $n$, and
  the size of the smallest $\uparrow$-RFTA which recognizes $L_n$
  is a linear function of $n$.
\end{proposition}

\begin{sketchofproof}
  We give an example of regular tree languages for which the size of
  the $\uparrow$-DFTA grows exponentially with respect to the size of
  the equivalent canonical $\uparrow$-RFTA.  A path is a sequence of
  symbols from the root to a leaf of a tree. The length
  of a path is the number of symbols on the path, except the root. Let
  $\F=\{f(,),a\}$ and let us consider the tree language $L_n$ which
  contains exactly the trees with at least one path of length $n$.
  Let $A_n=\aaa$ be a $\uparrow$-FTA defined by:
  $Q=\{q_{*},q_0,\hdots,q_n\},Q_f=\{q_0\}$ and
  \begin{multline*}
    \Delta=\{a \rightarrow q_{*}, a \rightarrow q_n, f(q_{*},q_{*})
    \rightarrow
    q_{*}\} \cup \\
    \bigcup_{k\in [1,\dots,n],q\in Q\setminus\{q_0\}}^n \big\{ f(q_k,q)
    \rightarrow q_{k-1}, f(q,q_{k}) \rightarrow q_{k-1}, f(q_k,q)
    \rightarrow q_{*}, f(q,q_{k}) \rightarrow q_{*} \big\}
  \end{multline*}
  
  Let $C_*$ be the set of contexts which contain at least one path
  of length $n$. Let $C_i$ be the set of contexts whose path from
  the root to $\diamond$ is of length $i$.  Let $t_*$ be a term such
  that all its paths are of length greater than $n$. Note that the
  set of contexts $c$ such that $c[t_*]$ belongs to $L_n$ is exactly
  the set of contexts $C_*$. Let $t_0 \hdots t_n$ be terms such that
  for all $i \leq n$, $t_i$ contains exactly one path of length
  smaller than $n$, and the length of this path is $n-i$. Therefore,
  $t_i^{-1}L_n$ is the set of contexts $C_* \cup C_i$.
  
  One can verify that $C_{q_*}$ is exactly $t_*^{-1}L_n=C_*$, and for
  all $i \leq n$, $C_{q_i}$ is exactly $t_i^{-1}L_n=C_* \cup C_i$.
  The reader should note that rules of the form $f(q_k,q) \rightarrow
  q_{*}$ and $f(q,q_{k}) \rightarrow q_{*}$ are not useful to
  recognize $L_n$ but they are required to obtain a $\uparrow$-RFTA
  (because $C_i$ is not a residual language of $L_n$). So $A_n$ is a
  $\uparrow$-RFTA and recognizes $L_n$. The size of $A_n$ is $n+2$.
  
  The construction of the smallest $\uparrow$-DFTA which recognizes
  $L(A_n)$ is left to the reader. But, it can easily be shown that the
  number of states is in $O(2^n)$ because states must store lengths of
  all paths smaller than $n$.
\end{sketchofproof}

Unfortunately, the size of a $\uparrow$-RFTA can be exponentially
larger than the size of an equivalent $\uparrow$-FTA.

\subsection{The canonical form of bottom-up residual tree automata}

As $\uparrow$-DFTA, $\uparrow$-RFTA have the interesting property to admit a 
canonical form. In the case of $\uparrow$-DFTA, there is a one-to-one
correspondence between residual languages and state languages. This is
a consequence of the Myhill-Nerode theorem for trees.

A similar result holds for $\uparrow$-RFTA. In a canonical
$\uparrow$-RFTA, the set of states is in one-to-one correspondence with a
subset of residual languages called prime residual languages.

\begin{definition}
  
  Let $L$ be a tree language. A bottom-up residual language of $L$ is
  \emph{composite} if and only if it is the union of the bottom-up
  residual languages that it strictly contains:
  $$t^{-1}L = \bigcup_{{t'}^{-1}L \subsetneq t^{-1}L} t'^{-1}L.$$
A residual language is \emph{prime} if and only if it is not composite.

\end{definition}

\begin{example}
  
  Let us consider again the tree languages in the proof of
  Proposition~\ref{prop:exporftadfta}. Let $Q_n$ be the set of states
  of $A_n$. All the $n+2$ states $q_*,q_0,\hdots,q_n$ of $Q_n$ have
  state languages which are prime residual languages. The subset
  construction applied on $A_n$ to build a $\uparrow$-DFTA $D_n$ leads
  to consider states which are subsets of $Q$. The state language of a
  state $\{q_{k_1} \hdots q_{k_n}\}$ is a composite residual language.
  It is the union of $t_{q_{k_1}}^{-1}L \hdots t_{q_{k_n}}^{-1}L$.

\end{example}       

In  canonical $\uparrow$-RFTAs, all state languages are prime
residual languages.

\begin{theorem} \label{theorem_can_asc}
  Let $L$ be a regular tree language and let us consider the
  $\uparrow$-FTA $A_{can} = \aaa$ defined by:

\begin{itemize}
\item $Q$ is in bijection with the set of all prime bottom-up residual
  languages of $L$. We denote by $t_q$ a ground term such that $q$ is
  associated with $t_q^{-1}L$ in this bijection
  
\item $Q_f$ is the set of all elements $q$ of $Q$ such that
  $t_q^{-1}L$ contains the void context $\diamond$,
  
\item $\Delta$ contains all the rules $f(q_1, \ldots, q_n) \rightarrow
  q$ such that $t_q^{-1}L \subseteq (f(t_{q_1}, \ldots,
  t_{q_n}))^{-1}L$ and all the rules $a \rightarrow q$ such that $a
  \in \F_0$ and $t_q^{-1}L \subseteq a^{-1}L$.
\end{itemize}
$A_{can}$ is a $\uparrow$-RFTA, it is the smallest $\uparrow$-RFTA
in number of states which recognizes $L$, and it is unique up to a
renaming of its states.
\end{theorem}

\begin{sketchofproof}
  There are three things to prove in this theorem: the canonical
  $\uparrow$-RFTA $A_{can}=\aaa$ of a regular tree language
  $L$ recognizes $L$, it is a $\uparrow$-RFTA, and there cannot be any
  strictly smaller $\uparrow$-RFTA which recognizes $L$. The three points are
  proved in this order.

  We first have to prove the equality $L(A_{can}) = L$. It follows from the
  identity $(\circledast)$ $\forall t,\ t^{-1}L=\bigcup_{q \in Q, \; t \lrproof{A_{can}} q} t_q^{-1}L$
  which can be proved inductively on the height of $t$. Using this
  property, we have: 

$$ t\in L \Leftrightarrow \diamond \in t^{-1}L 
\mathrel{\mathop{\kern 0pt{\Leftrightarrow}}\limits_{\circledast}} 
\diamond \in \bigcup_{q \in Q, \; t \lrproof{A_{can}} q} t_q^{-1}L
\Leftrightarrow \exists q_f \in Q_f, t \lrproof {A_{can}} q_f
\Leftrightarrow t \in L(A_{can})$$

  The equality between $L$ and $L(A_{can})$ helps us to prove the
  characterization of $\uparrow$-RFTA: $t_q^{-1}L=C_q^{A_{can}}$ 
 where $C^{A_{can}}_q$ is the state language of $q$ in $A_{can}$.


  
  
  The last point can be proved in such a way. In a $\uparrow$-RFTA,
  any residual language is a union of state languages, and any state
  language is a residual language. So any prime residual language is a
  state language, so there is at least as much states in a
  $\uparrow$-RFTA as prime residual languages admitted by its corresponding
  tree language.

\end{sketchofproof}  

The canonical automaton is uniquely defined determined by the tree
language under consideration, but there may be other automata which
have the same number of states. The canonical $\uparrow$-RFTA is
unique because it has the maximum number of rules.  Even though all
its states are associated to prime residual languages, the automaton
considered in the proof of Proposition~\ref{prop:exporftadfta} is not
the canonical one because some rules are missing: $\bigcup_{k=1}^{n} \{
f(q_k,q_0) \rightarrow q_{k-1}, f(q_0,q_k) \rightarrow q_{k-1} \}$ and
$\bigcup_{q \in Q} \{ f(q,q_0) \rightarrow q_{*}, f(q,q_0) \rightarrow
q_{*} \}$.

\section{Top-Down residual finite tree automata}
\label{sec:topdown}

The definition of top-down residual finite tree automata
($\downarrow$-RFTA) is tightly correlated with the definition of
$\uparrow$-RFTA. Similarly to $\uparrow$-RFTA, $\downarrow$-RFTA are
defined as non-deterministic tree automata where each state language
is a residual language.  Any $\downarrow$-RFTA can be transformed in a
canonical equivalent $\downarrow$-RFTA --- minimal in the number of states and
unique up to state renaming.

The main difference between the bottom-up and the top-down case is in
the problem of the expressive power of tree automata. The
three classes of bottom-up tree automata, $\uparrow$-DFTA, $\uparrow$-RFTA or
$\uparrow$-FTA, have the same expressive power. In the top-down case,
deterministic, residual and non-deterministic tree automata have
different expressive power. This makes the canonical form of
$\downarrow$-RFTA more interesting. Compared to the minimal form of
$\downarrow$-DFTA, it can be smaller when both exist, and it exists
for a wider class of tree languages.

Let us introduce $\downarrow$-RFTA through their similarity with
$\uparrow$-RFTA, then study this specific problem of expressiveness.

\subsection{Analogy with bottom-up residual tree automata}

Let us formally define state languages in the top-down case:

\begin{definition}
  Let $L$ be a regular tree language over a ranked alphabet $\F$, let
  $A$ be a top-down tree automaton
  which recognizes $L$, and let $q$ be a state of this automaton. The
  state language of $L$ relative to $q$, written $L_q$, is the set of
  terms which are accepted by $q$:
  
  $$L_q = \{ t \in \T(\F) \mid q(t) \lrproof{A} t \}.$$

\end{definition}

It follows from this definition some properties similar to those
already studied in the previous section.  Firstly, state languages are
generally not residual languages.  Secondly, residual languages are
unions of state languages. Let us define $Q_c$:

$$Q_c = \{ q \mid q \in Q, \exists q_i \in I, q_i ( c [ \diamond ] )
\rightarrow_A^* c [ q( \diamond ) ] \}.$$

We have the following relation between state languages and
residual languages.

\begin{lemma}
  \label{cL_union_des_Lq}
  
  Let $L$ be a tree language and let $A = \ada$ be a top-down tree
  automaton which recognizes $L$. Then $\forall c \in \C(\F),
  \bigcup_{q \in Q_c} L_q = c^{-1} L.$

\end{lemma}

These similarities lead us to this definition of top-down residual
tree automata:

\begin{definition}
  A top-down Residual Finite Tree Automaton ($\downarrow$-RFTA) 
  recognizing a tree language $L$ is a $\downarrow$-FTA $
  A = \ada$  such that: $\forall
  q \in Q$, $\exists c \in \CTX$, $L_q = c^{-1} L.$
\end{definition}

Languages defined in the proof of Proposition~\ref{prop:exporftadfta}
are still interesting here to define examples of top-down residual
tree automata:

\begin{example}
\label{exemple_alain_mirroir}

Let us consider again the family of tree languages $L_n$, and the
family of corresponding $\uparrow$-RFTA $A_n$.
For every $n$, let $A'_n$ be the $\downarrow$-RFTA defined by:
$Q=\{q_{*},q_0,\hdots,q_n\},Q_i=\{q_0\}$ and $\Delta=\{q_{*}(a)
\rightarrow a, q_n(a) \rightarrow a, q_{*}(f) \rightarrow
f(q_{*},q_{*})\} \cup \bigcup_{k=1}^n \{ q_{k-1}(f) \rightarrow
f(q_k,q_{*}) , q_{k-1}(f) \rightarrow f(q_{*},q_k) \}$.
  
For every $k \leq n$, the state language of $q_{k}$ is equal to
$L_{n-k}$. And, $L_{n-k}$ is the top-down residual language of $c_k$,
where $c_k$ is a context whose height from the root to the special
constant $\diamond$ is $k$ and $c_k$ does not contain any path whose
length is smaller or equal to $n$.  The state language of $q_{*}$ is
$\T(\F)$. And, $\T(\F)$ is the top-down residual language of $L_n$
relative to $c_*$, where $c_*$ is a context who contains a path
whose length is $n$. So $A'_n$ is a $\downarrow$-RFTA. Moreover, it is
easy to verify that $A'_n$ recognizes $L_n$.

\end{example}

\subsection{The expressive power of top-down tree automata}

\paragraph{Top-down deterministic automata and path-closed languages}

A tree language $L$ is
\emph{path-closed} if:
$$\forall c \in C(\F), c[f(t_1,t_2)] \in L \wedge c[f(t'_1,t'_2)] \in
L \Rightarrow c[f(t_1,t'_2)] \in L.$$

The reader should note that the definition only considers binary
symbols, the definition can easily be extended to $n$-ary symbols. The
class of languages that $\downarrow$-DFTA can recognize is the class
of path-closed languages~\cite{Viragh81}.


\paragraph{Context-deterministic automata and homogeneous languages.}

Podelski and Nivat in~\cite{NivatPodelski97} have defined
\emph{l-r-deterministic} top-down tree automata. In the present paper,
let us call them top-down \emph{context-deterministic} tree automata.

\begin{definition}
  A top-down context-deterministic tree automaton ($\downarrow$-CFTA)
  $A$ is a $\downarrow$-FTA such that for every context $c \in
  \C(\F)$, $Q_c$ is either the empty set or a singleton set.
\end{definition}

An \emph{homogeneous language} is a tree language $L$ satisfying:

$$\forall c \in C(\F), c[f(t_1,t_2)] \in L \wedge c[f(t_1,t'_2)] \in L
\wedge c[f(t'_1,t_2)] \Rightarrow c[f(t'_1,t'_2)] \in L.$$

Again, the definition can easily be extended from the binary case to
$n$-ary symbols. They have shown that the class of languages recognized
by $\downarrow$-CFTA is the class of homogeneous languages.

\paragraph{The hierarchy}

A $\downarrow$-DFTA is a $\downarrow$-CFTA. For $\downarrow$-CFTA and
$\downarrow$-RFTA, we have the following result: 

\begin{lemma}
  Any trimmed  $\downarrow$-CFTA is a $\downarrow$-RFTA.
\end{lemma}

\begin{proof}
  Let $A=\ada$ be a trimmed $\downarrow$-CFTA recognizing a tree
  language $L$. As $A$ is trimmed, all states are reachable, so for
  every $q$, there exists a $c$ such that $q \in Q_c$. Then, by
  definition of a $\downarrow$-CFTA, for every $q$, there exists a $c$
  such that $\{q\}=Q_c$. Using Lemma~\ref{cL_union_des_Lq}, we have:

  $$ \forall q \in Q, \exists c \in \C(\F), L_q = c^{-1} L.$$ 

  stating that $A$ is a $\downarrow$-RFTA.
\qed \end{proof}

Therefore, if we denote by $\mathcal{L}_{\mathcal{C}} $ the class
of tree languages recognized by a class of automata $\mathcal{C}$, we
obtain the following hierarchy:

$$\mathcal{L}_{\downarrow-DFTA} \subseteq
\mathcal{L}_{\downarrow-CFTA} \subseteq \mathcal{L}_{\downarrow-RFTA}
\subseteq \mathcal{L}_{\downarrow-FTA} $$

\paragraph{The hierarchy is strict}

\begin{itemize}
\item Let $L=\{f(a,b),f(b,a)\}$. $L_1$ is homogeneous but not
  path-closed. Therefore $L$ can be recognized by a
  $\downarrow$-CFTA, but can not be recognized by a $\downarrow$-DFTA.

\item The tree languages $L_n$ in the proof of
Proposition~\ref{prop:exporftadfta} are not recognized by 
$\downarrow$-CFTA.  We can easily verify that $L_n$ is not
homogeneous. Indeed, if $t$ is a term which has a path whose length
is equal to $n-1$, and $t'$ a term which does not have any path
whose length is smaller than $n$, $f(t,t)$, $f(t,t')$, $f(t',t)$
belong to $L_n$, but $f(t',t')$ does not. And, we have already shown
that $L_n$ is recognized by a $\downarrow$-RFTA.

\item Let $L'=\{f(a,b),f(a,c),f(b,a),f(b,c),f(c,a),f(c,b)\}$.  $L'$
  is a finite language, therefore it is a regular tree language which
  can be recognized by a $\downarrow$-FTA. $L'$
  cannot be recognized by a $\downarrow$-RFTA. To
  prove that, let us consider $A'$ a $\downarrow$-FTA which
  recognizes $L'$. The top-down residual languages of $L'$ are
  $\{a,b\}$, $\{a,c\}$, $\{b,c\}$ and $L'$. As $A'$ recognizes
  $L'$, it recognizes $f(a,b)$.  This implies the existence of three
  states $q_1$, $q_2$, $q_3$ and three rules $q_1(f) \rightarrow
  f(q_2,q_3)$, $q_2(a) \rightarrow a$, and $q_3(b) \rightarrow b$. If
  $A'$ was a $\downarrow$-RFTA, then $q_2$ would accept a residual
  language. As $q_2$ accepts $a$, it would accept either $\{a,b\}$ or
  $\{a,c\}$. Similarly, $q_3$ would accept either $\{a,b\}$ or
  $\{b,c\}$. In these conditions, and thanks to the rule $q_1(f)
  \rightarrow f(q_2,q_3)$, $A'$ would recognize $f(a,a)$, $f(b,b)$ or
  $f(c,c)$. So $A'$ cannot be a $\downarrow$-RFTA.

\end{itemize}

Therefore, we obtain the following result:

\begin{theorem}
  $\mathcal{L}_{\downarrow-DFTA} \subsetneq
  \mathcal{L}_{\downarrow-CFTA} \subsetneq
  \mathcal{L}_{\downarrow-RFTA} \subsetneq
  \mathcal{L}_{\downarrow-FTA} $
\end{theorem}

So top-down residual tree automata are strictly more expressive than
context-deterministic tree automata. But as far as we know, there is
no straightforward characterization of the tree languages recognized by
$\downarrow$-RFTA.

\subsection{The canonical form of top-down residual tree automata}

The problem of the canonical form of top-down tree automata is similar
to the bottom-up case. Whereas there is no way to reduce a
non-deterministic top-down tree automaton to a unique canonical form,
a top-down residual tree automaton can take such a form. Its
definition is similar to the definition of the canonical bottom-up
tree automaton.

In the same way that we have defined composite
bottom-up residual language, a  top-down residual language of $L$ 
is \emph{composite} if and only if it is the union of the top-down 
residual languages that it strictly contains and a residual language 
is \emph{prime} if and only if it is not composite.

\begin{theorem}
\label{theorem_can_des}
  
Let $L$ be a tree language in the class
$\mathcal{L}_{\downarrow-RFTA}$. Let us consider the $\downarrow$-RFTA 
$A_{can}=\fta(Q,\F,I,\Delta)$ defined by:

  \begin{itemize}
    
  \item $Q$ is a set of state in bijection with the prime residual
    languages of $L$. For each of these residual languages, there
    exists a $c_q$ such that $q$ is associated with $c_q^{-1}L$ in this
    bijection.
    
  \item $I$ is the set of prime residuals which are subsets of $L$.
    
  \item $\Delta$ contains all the rules $q(a) \rightarrow a$ such that
    $a$ is a constant and $c_q[a] \in L$, and all the rules $q(f)
    \rightarrow f(q_1, \hdots, q_n)$ such that for all $t_1 \hdots
    t_n$ where $t_i \in c_{q_i}^{-1}L$, $c_q[f(t_1,\hdots,t_n)] \in
    L$.

  \end{itemize}
  
  $A_{can}$ is a $\downarrow$-RFTA, it is the smallest
  $\downarrow$-RFTA in number of states which recognizes $L$, and it
  is unique up to a renaming of its states.
\end{theorem}

\begin{sketchofproof}

The proof is mainly based on this lemma: $t \in c_q^{-1}L \Leftrightarrow t \in L_q^{A_{can}}$

where $L_q^{A_{can}}$ is the state language of $q$ in the automaton
$A_{can}$.

This lemma is proved by induction on the height of $t$. This is not a
straightforward induction. It involves the rules of a
$\downarrow$-RFTA automaton $A'$ which recognizes $L$. Its existence is
granted by the hypothesis of the theorem.

Once this is proved, it can be easily deduced that $A_{can}$
recognizes $L$ and is a RFTA. As there is one state per prime residual
in $A_{can}$, it is minimal in number of states.

\end{sketchofproof}

\section{Decidability issues}

Some decision problems naturally arise with the definition of RFTA. Most of these problems are solved just noting that one can build all residual languages of a given regular language $L$ defined by a non-deterministic tree automaton. In the bottom-up case, the state languages of the minimal $\uparrow$-RFTA which recognizes $L$ are exactly the residual languages of $L$, and this automaton can be built with the subset construction. In the top-down case, the subset construction does not necessarily gives us an automaton which recognizes exactly $L$, but it gives us the set of all residual languages. Therefore, knowing whether a tree automaton is a RFTA, whether a residual language is prime or composite, and whether a tree automaton is a canonical RFTA are decidable. These problems have not been deeply studied in terms of complexity, but they are at least as hard as the similar problems with strings, that is they are PSPACE-hard (\cite{DenisLemayTerlutte2002a}).

\section{Conclusion}
We have defined new classes of non-deterministic tree automata. In the
bottom-up case, we get another characterization of regular tree
languages. More interestingly, in the top-down case, we obtain a
subclass of the regular tree languages. For both cases, we have a
canonical form and the size of residual tree automata can be much
smaller than equivalent (when exist) deterministic ones.

We are currently extending these results to the case of unranked trees
because our application domain is concerned with html and xml
documents. Also, we are designing learning algorithms for residual finite
tree automata extending previous algorithms for residual finite
string automata~\cite{DenisLemayTerlutte2001c,DenisLemayTerlutte2002b}.

\newcommand{\etalchar}[1]{$^{#1}$}

\newpage
\appendix
\section{Appendix}

\subsection{Proof of Equation \eqref{eq:1}}
\label{sec:proof-equat-eqref}

Let $L$ be a tree language and $\aaa$ a $\uparrow$-FTA which
recognizes it. We show that $\forall t \in \TF, t^{-1}L = \bigcup_{t
  \lrproof{A} q} C_q.$

Let $t \in \T(\F)$, and $c \in t^{-1}L$. $c[t] \in L$, so there exists
$q_f \in Q_f$ and $q \in Q$ such that $c[t] \lrproof{A} c[q]
\lrproof{A} q_f$, where $t \lrproof{A} q$ and $c \in C_q$. So $c \in
\bigcup_{t \lrproof{A} q} C_q$. So $t^{-1}L \subseteq \bigcup_{t
  \lrproof{A} q} C_q$

Let $t \in \T(\F)$, and $c \in \bigcup_{t \lrproof{A} q} C_q$. There
exists a $q \in Q$ such that $t \lrproof{A} q$ and $c \in C_q$. So
there exists $q_f \in Q_f$ such that $c[t] \lrproof{A} c[q]
\lrproof{A} q_f$. So $c \in t^{-1}L$. So $\bigcup_{t \lrproof{A} q}
C_q \subseteq t^{-1}L$

\subsection{Proof of the theorem \ref{theorem_can_asc}}

\begin{theorem}

The canonical $\uparrow$-RFTA recognizing a regular tree language is the smallest $\uparrow$-RFTA which recognizes it. Therefore, $\uparrow$-RFTA accepts a unique and minimal representation.

\end{theorem}

The first point we have to demonstrate in this theorem is that the canonical $\uparrow$-RFTA that we have defined recognizes $L$. 

Before this demonstration, we need to establish two properties of residual languages:

\begin{lemma}
\label{can_lemme1}

Let $L$ a regular language. 
$$\forall i, 1 \leq i \leq n, t_i^{-1}L \subseteq t'^{-1}_i L \Rightarrow f(t_1,\hdots,t_n)^{-1}L \subseteq f(t'_1,\hdots,t'_n)^{-1}L$$

\end{lemma}

\begin{proof}
  
  This lemma can be proven inductively on $i$. Let $t_1 \hdots t_n$
  such that for all $i$, $t_i^{-1}L$ is a subset of $t'^{-1}_i L$. Let
  $c$ in $f(t_1,\hdots,t_n)^{-1}L$.  Let us assume that
  $c[f(t'_1,\hdots,t'_{i-1},t_i,\hdots,t_n)] \in L$. This implies that
  $c[f(t'_1,\hdots,t'_{i-1},\diamond,t_{i+1},\hdots,t_n)] \in t_i^{-1}
  L$, and therefore
  $c[f(t'_1,\hdots,t'_{i-1},\diamond,t_{i+1},\hdots,t_n)] \in
  t'^{-1}_i L$.  
  
  So $c[f(t'_1,\hdots,t'_i,t_{i+1},\hdots,t_n)] \in L$.  Inductively,
  $c[f(t'_1,\hdots,t'_n)] \in L$.  
  
  So $f(t_1,\hdots,t_n)^{-1}L
  \subseteq f(t'_1,\hdots,t'_n)^{-1}L$.

\qed \end{proof}

\begin{lemma}
\label{can_lemme2}

$$ \forall i, 1 \leq i \leq n, t_i^{-1}L = \bigcup_{j_i} t_{i,j_i}^{-1}L \Rightarrow f(t_1,\hdots,t_n)^{-1}L = \bigcup_{j_1 \hdots j_n} f(t_{1,j_1},\hdots,t_{n,j_n})^{-1}L$$

Here, $\bigcup_{j_1 \hdots j_n}$ has to be understood as 'the union of all the possible combination of $j_1 \hdots j_n$'. 

\end{lemma}

\begin{proof}

Let $t_1 \hdots t_n$ and for all $i \leq n$, $t_{i,1} \hdots t_{i,m_i}$ such that $t_i^{-1}L = \bigcup_{1 \leq j_i \leq m_i} t_{i,j_i}^{-1}L$.

$$\forall t_{1,j_1} \hdots t_{n,j_n}, \forall i \leq n, t_{i,j_i}^{-1}L \subseteq t_i^{-1}L  \Rightarrow_{lemma \ref{can_lemme1}} $$
$$\forall t_{1,j_1} \hdots t_{n,j_n}, f(t_{1,j_1} \hdots t_{n,j_n})^{-1}L \subseteq f(t_1,\hdots,t_n)^{-1}L \Rightarrow$$ 
$$\bigcup_{j_1 \hdots j_n} f(t_{1,j_1},\hdots,t_{n,j_n})^{-1}L \subseteq  f(t_1,\hdots,t_n)^{-1}L$$

Now, let $c$ in $f(t_1,\hdots,t_n)^{-1}L$.
$$c[f(t_1,\hdots,t_n)] \in L \Rightarrow c[f(\diamond,t_2,\hdots,t_n)] \in t_1^{-1}L$$

As $t_1^{-1}L =  \bigcup t_{1,j}^{-1}L$, there exists $t_{1,m_1}$ such that $c[f(\diamond,t_2,\hdots,t_n)] \in t_{1,m_1}^{-1}L$. So $c[f(t_{1,m_1},t_2,\hdots,t_n)] \in L$.

It can be proven inductively on $i$ that there exists $t_{1,m_1} \hdots t_{n,m_n}$ such that $c[f(t_{1,k_1},\hdots,t_{n,m_n})] \in L$. So $c \in f(t_{1,m_1},\hdots,t_{n,m_n})^{-1}L$. So:

$$ f(t_1,\hdots,t_n)^{-1}L \subseteq \bigcup_{j_1 \hdots j_n} f(t_{1,j_1},\hdots,t_{n,j_n})^{-1}L$$

\end{proof} \qed

Now, we can prove inductively this lemma, which is the main step to prove the equality between $L$ and $L(A_{can})$

\begin{lemma}
\label{can_lemme3}

$$\forall t, t^{-1}L = \bigcup_{t \lrproof{A_{can}} q} t_q^{-1}L$$

\end{lemma}

\begin{proof}

Let us prove this lemma inductively. Let $h(t)$ be the height of $t$.

Let us assume that $h(t)=1$, so $t=a$ where $a$ is a constant. A residual is a union of prime residuals, so:
$$a^{-1}L = \bigcup_{t_q^{-1}L \subseteq a^{-1}L} t_q^{-1}L$$

As $t_q^{-1}L \subseteq a^{-1}L$ if and only if $A_{can}$ contains the rule $a \rightarrow q$:
$$ a^{-1}L= \bigcup_{a \lrproof{A_{can}} q} t_q^{-1}L$$

Now let us assume that for any term $t$ such that $h(t) \leq k$, lemma \ref{can_lemme3} is true.

Let $t=f(t_1,\hdots,t_n)$ such that $h(t)=k+1$. 

$$h(t)=k+1 \Rightarrow \forall i \leq n, h(t_i)=k \Rightarrow \forall i, t_i^{-1}L=\bigcup_{t_i \lrproof{A_{can}} q_{i,j_i}} t_{ q_{i,j_i}}^{-1}L \Rightarrow_{lemma \ref{can_lemme2}}$$
$$ t^{-1}L=\bigcup_{t_i \lrproof{A_{can}} q_{i,j_i}} f(t_{1,q_1},\hdots,t_{n,q_n})^{-1}L$$

Any residual is a union of prime residuals, so for all $j_1 \hdots j_n$:
$$f(t_{q_{1,j_1}},\hdots,t_{q_{n,j_n}})^{-1}L = \bigcup_{t_q^{-1}L \subseteq f(t_{q_{1,j_1}},\hdots,t_{q_{n,j_n}})^{-1}L} t_q^{-1}L$$

So:
$$t^{-1}L= \bigcup_{t_i \lrproof{A_{can}} q_{i,j_i}} f(t_{1,q_1},\hdots,t_{n,q_n})^{-1}L \Rightarrow$$
$$t^{-1}L= \bigcup_{t_i \lrproof{A_{can}} q_{i,j_i}} ( \bigcup_{t_q^{-1}L \subseteq f(t_{q_{1,j_1}},\hdots,t_{q_{n,j_n}})^{-1}L} t_q^{-1}L)$$

As $t_q^{-1}L \subseteq f(t_{q_{1,j_1}},\hdots,t_{q_{n,j_n}})^{-1}L$ if and only if $A_{can}$ contains the rule $f(q_{1,j_1},\hdots,q_{n,j_n}) \rightarrow q$,

$$t^{-1}L=\bigcup_{t \lrproof{A_{can}} q} t_q^{-1}L$$

\end{proof} \qed

The equality between $L$ and $L(A_{can})$ is formalized as such:

\begin{lemma}

The canonical $\uparrow$-RFTA $\A_{can}$ of a language $L$ recognizes $L$, that is:

$$\exists q_f \in Q_f, t \lrproof {A_{can}} q_f \Leftrightarrow t \in L$$

\end{lemma}

\begin{proof}

Let $t \in \T(\F)$ and $q_f \in Q_f$ such that $t \lrproof{A_{can}} q_f$.

$$t \lrproof{A_{can}} q_f \Rightarrow t_{q_f}^{-1}L \subset \bigcup_{t \lrproof{A_{can}} q_j} t_{q_j}^{-1}L \Rightarrow_{lemma \ref{can_lemme3}} t_{q_f}^{-1}L \subset t^{-1}L$$

As $\diamond \in t_{q_f}^{-1}L$, $\diamond \in t^{-1}L$, so $t \in L$.

Let $t \in L$.

$$\diamond \in t^{-1}L \Rightarrow \exists q_j \mid t \lrproof{A_{can}} q_j \wedge \diamond \in t_{q_j}^{-1}L \Rightarrow$$
$$ \exists q_j \mid t \lrproof{A_{can}} q_j \wedge q_j \in Q_f$$

\end{proof}

Now, we have to prove that the canonical $\uparrow$-RFTA is a $\uparrow$-RFTA. In order to do this, we need to establish this lemma:

\begin{lemma}
\label{can_lemme4}

Let $t_q^{-1}L$ and $t_{q'}^{-1}L$ be prime bottom-up residual languages of $L$. Let $C_q^{A_{can}}$ and $C_{q'}^{A_{can}}$ be sets of contexts accepted by $q$ and $q'$ in then canonical automaton of $L$ $A_{can}$. Then:

$$t_{q'}^{-1}L \subset t_q^{-1}L \Rightarrow C^{A_{can}}_{q'} \subset C^{A_{can}}_q$$

\end{lemma}

\begin{proof}
Let $t_q$ and $t_{q'}$ such that $t_{q'}^{-1} L \subset t_q^{-1} L$. For all $t_{q_1} \hdots t_{q_n}$, $f(t_{q_1},\hdots,t_{q'},\hdots,t_{q_n})^{-1} L \subset f(t_{q_1},\hdots,t_q,\hdots,t_{q_n})$ (lemma \ref{can_lemme1}).

The construction of the set of rules of the canonical automaton implies that:
$$ f(q_1,\hdots,q_n) \rightarrow {q'} \in \Delta \Leftrightarrow t_{q'}^{-1}L \subseteq f(t_{q_1},\hdots,t_{q_n})^{-1}L$$

So:
$$ f(q_1,\hdots,q',\hdots,q_n) \rightarrow q'' \in \Delta \Rightarrow$$
$$ t_{q''}^{-1}L \in f(t_{q_1},\hdots,t_{q'},\hdots,t_{q_n})^{-1}L \Rightarrow$$
$$ t_{q''}^{-1}L \in f(t_{q_1},\hdots,t_q,\hdots,t_{q_n})^{-1}L \Rightarrow$$
$$ f(q_1,\hdots,q,\hdots,q_n) \rightarrow q'' \in \Delta$$

So each context accepted by $q'$ is accepted by $q$.

So $C^{A_{can}}_{q'} \subset C^{A_{can}}_q$.
\qed \end{proof}

\begin{lemma}

The canonical RFTA $A_{can}$ of a language $L$ is a residual finite tree automata.

\end{lemma}

\begin{proof}

Let $t_q^{-1}L$ be a prime residual language of $L$. Thanks to lemma \ref{can_lemme3}:

$$t_q^{-1}L=\bigcup_{t_q \lrproof{A_{can}} q'} t_{q'}^{-1}L$$

If  $t_q^{-1}L$ would strictly contain all the $t_{q'}^{-1}L$ of the union, it would be composite. As it is prime, $t_q^{-1}L$ is itself an element of this union, so $t_q \lrproof{A_{can}} q$.

Equation \eqref{eq:1} tells us that:

$$t_q^{-1}L = \bigcup_{t_q \lrproof{A_{can}} q'} C_{q'}$$

So $C_q \subset t_q^{-1}L$.

For all $q'$ such that $t_q \lrproof{A_{can}} q'$, $t_{q'}^{-1}L \subset t_q^{-1}L$, so $C_{q'} \subset C_q$ (lemma \ref{can_lemme4}). As the union of all $C_{q'}$ is equal to $t_q^{-1}L$, $t_q^{-1}L \subset C_q$  

So $t_q^{-1}L = C_q$, so every prime residual language is accepted by its corresponding state.

So $A_{can}$ is a RFTA. 

\qed \end{proof}

\begin{lemma}
  
  The canonical RFTA $A_{can}$ of a language $L$ is the smallest RFTA
  which recognizes $L$.

\end{lemma}

\begin{proof}

Let $A_{can}$ be the canonical RFTA of a language $L$, and $t$ such that $\nexists q \in Q, t^{-1} L=C_q$.

Thanks to lemma \ref{can_lemme3}, $t^{-1} L = \bigcup_{t \lrproof{A_{can}} q} t_q^{-1} L$. As  $\nexists q \in Q, t^{-1} L=t_q^{-1}L$, $t^{-1} L$ is a union of residuals that it strictly contains. So $t^{-1}L$ is a composite residual.

So for all prime residuals $t^{-1}L$, there is a $q$ such that $t^{-1}L=C_q$. $A_{can}$ contains as much states as prime residuals in $L$, so it is the smallest RFTA which recognizes $L$. 

\qed \end{proof}

\subsection{Proof of the theorem \ref{theorem_can_des}}

\begin{theorem}

Let $L$ be a language recognized by a $\downarrow$-RFTA. The canonical top-down residual tree automaton of $L$ is the smallest $\downarrow$-RFTA which recognizes $L$.

\end{theorem}

In order to prove this theorem, let us firstly prove these lemma:

\begin{lemma}
\label{cand_lemme1}

Let $A=\fta(Q,\F,I,\Delta)$ be a $\downarrow$-RFTA which recognizes $L$. For any prime residual $c^{-1}L$, there exists a state $q \in Q$ such that $L_q=c^{-1}L$.

\end{lemma}

\begin{proof}

Let $c$ be a context of $L$ such that $\nexists q \in Q, c^{-1}L=L_q$. 

 Lemma \ref{cL_union_des_Lq} implies that $c^{-1}L=\bigcup_{q \in Q_c} L_q$ and none of these $L_q$ are equal to $c^{-1}L$. As $\forall q \in Q, L_q=c_q^{-1}L$, we have $c^{-1}L=\bigcup_{q \in Q_c} c_q^{-1}L$ where none of the $c_q^{-1}L$ are equal to $c^{-1}L$. So $c^{-1}L$ is composite.

\qed \end{proof} 

Now, let us make the main part of the demonstration: let us prove that each prime residual language is exactly accepted by a state of the canonical $\downarrow$-RFTA.

\begin{lemma}
\label{cand_lemme2}

Let $L$ be a language recognized by a $\downarrow$-RFTA. Let $A_{can}=\fta(Q,\F,I,\Delta)$ be its canonical automaton. For all $q$ in $Q$, $c_q^{-1}L = L_q$.

\end{lemma}

\begin{proof} 

As seen in the definition, $Q$ is in bijection with the set of all residual languages, so for all $q$ there exists a corresponding $c_q^{-1}L$.  Let us prove inductively on the height of $t$ that $t \in c_q^{-1}L \Leftrightarrow t \in L_{A_{can},q}$. Let us call $H(n)$ this hypothesis when $h(t) \leq n$.\\


Firstly, let us prove $H(1)$.

Let $t$ such that $h(t)=1$ and $t \in c_q^{-1}L$. As $h(t)=1$, $t=a$ where $a$ is a constant. As $t \in c_q^{-1}L$, $c_q[a] \in L$. So $\Delta$ contains the rule $q(a) \rightarrow a$, so $t \in L_{A_{can},q}$. Reciprocally, $t \in L_{A_{can},q}$ where $t=a$ implies that $\Delta$ contains the rule $q(a) \rightarrow a$. This rule exists in the canonical automata if and only if $a$ is a constant and $c_q[a] \in L$. So $c_q[a] \in L$, so $t \in c_q^{-1}L$.\\

Now, let us assume that $H(l)$ is true when $l<k$. Let us prove that $H(k)$ is true.

Let $t=f(t_1,\hdots,t_n) \in c_q^{-1}L$ such that $h(t)=k$. For all $t_i$ where $1 \leq i \leq n$, $t_i \in c_q[f(t_1,\hdots,t_{i-1},\diamond,t_{i+1},\hdots,t_n)]^{-1}L$.

Now, let us consider $A'=\fta(Q',\F,I',\Delta')$ a $\downarrow$-RFTA which recognizes $L$. As $L$ is recognized by a $\downarrow$-RFTA, $A'$ exists. We will use this automaton to prove the existence of a rule  $q \rightarrow f(q_1,\hdots,q_n)$ such that for all $i$, $q_i[t_i] \lrproof{A} t_i$ in $A_{can}$.

As $c_q^{-1}L$ is prime, there exists a $q' \in Q'$ such that $L_{A',q'}=c_q^{-1}L$ (lemma \ref{cand_lemme1}). As $t \in c_q^{-1}L$, there exists in $\Delta'$ a rule $q' \rightarrow f(q'_1,\hdots,q'_n)$ such that for all $i$, $1 \leq i \leq n$, we have  $t_i \in L_{A',q'_i}$. 

For all $t'_1 \hdots t'_n$ such that $t'_i \in L_{A',q'_i}$, $f(t'_1,\hdots,t'_n) \in c_q^{-1}L$.

As $L_{A',q'_i}$ is a residual, it is either a prime residual or a composite residual. If it is a prime residual, there exists a $q_i \in Q$ such that $L_{A',q'_i}=c_{q_i}^{-1}L$ and $t_i \in c_{q_i}^{-1}L$. If it is a composite residual, there exists a $q_i \in Q$ such that $c_{q_i}^{-1}L \subset L_{A',q'_i}$ and $t_i \in c_{q_i}^{-1}L$.

So there exists $q_1 \hdots q_n$ such that $t_i \in c_{q_i}^{-1}L \subset L_{A',q'_i}$. So for all $t'_1 \hdots t'_n$ in $c_{q_1}^{-1}L \hdots c_{q_n}^{-1}L$, $f(t'_1,\hdots,t'_n) \in L_{A',q'}^{-1}L=c_q^{-1}L$. So the rule $q(f) \rightarrow f(q_1,\hdots,q_n)$ exists in $\Delta$. 

For all $t_i$, $h(t_i)<k$, so as we have assumed that $H(l)$ is right when $l<k$, $H(h(t_i))$ is right. So for all $i$, $t_i \in L_{A_{can},q_i}$. As $q(f) \rightarrow f(q_1,\hdots,q_n)$, $t \in L_{{A_{can}},q}$.

We have proven that $t \in c_q^{-1}L \Rightarrow t \in L_{A_{can},q}$. Now let us prove that $t \in L_{A_{can},q} \Rightarrow t \in c_q^{-1}L$.\\

Let $t=f(t_1,\hdots,t_n) \in L_{A_{can},q}$ such that $h(t)=k$.

There exist $q_1 \hdots q_t$ such that $q(f(t_1 \hdots t_n)) \lrproof{A} f(q_1(t_1),\hdots,q_n(t_n)) \lrproof{A} f(t_1,\hdots,t_n)$. For all $i$, $t_i \in L_{A_{can},q_i}$ and $h(t_i)<k$, so $H(h(t_i))$ is assumed to be true, so $t_i \in c_{q_i}^{-1}L$. The existence of the rule $q(f) \rightarrow f(q_1,\hdots,q_n)$ in $\Delta$ implies that for all $t'_1 \hdots t'_n$ such that $t'_i \in c_{q_i}^{-1}L$, $c_q[f(t'_1, \hdots, t'_n)] \in L$. So $t \in c_q^{-1}L$.

So H(k) is true. We have proven inductively that for any $t$, $t \in L_{A_{can},q} \Leftrightarrow t \in c_q^{-1}L$. 

\qed \end{proof}

\begin{lemma}

$A_{can}=<Q,\F,Q_i,\Delta>$ is a $\downarrow$-RFTA, recognizes $L$, and is minimal in number of states.

\end{lemma} 

\begin{proof}

Let us prove that lemma \ref{cand_lemme2} implies that $L(A_{can})=L$. Let $t \in L$. $\diamond^{-1}L=L$ is a residual, so it is a union of prime residuals. So there exists $q_i \in Q$ such that $t \in c_{q_i}^{-1}L$ and $c_{q_i}^{-1}L \subseteq L$. As $c_{q_i}^{-1}L = L_{A_{can},q_i}$, we have $t \in L_{A_{can},q_i}$. $c_{q_i}^{-1}L \subseteq L$, so $q_i$ is initial, so $t \in L(A_{can})$.

Reciprocally, let $t \in L(A_{can})$. There exists a $q_i \in I$ such that $t \in L_{A_{can},q_i}$. $c_{q_i}^{-1}L = L_{q_i}$, so $t \in c_{q_i}^{-1}L$. As $q_i$ is initial, $c_{q_i}^{-1}L$ is a subset of $L$. So $t \in L$. So $L=L(A_{can})$.

So $A_{can}$ recognizes $L$. For any $q$, $L_q = c_q^{-1}L$, so $A_{can}$ is a RFTA. For any prime residual of $L$, there exists a state in the RFTA which recognizes it. As there are one state per prime residual in $A_{can}$, $A_{can}$ is minimal in number of states.

\qed \end{proof}

\end{document}